# Single-Molecule Magnets: High-Field Electron Paramagnetic Resonance Evaluation of the Single-Ion Zero-Field Interaction in a $Zn^{II}_3Ni^{II}$ Complex

En-Che Yang,[†] Cem Kirman,[‡] Jon Lawrence,[‡] Lev N. Zakharov,[†] Arnold L. Rheingold,[†] Stephen Hill,[‡]*
and David N. Hendrickson[†]*

Department of Physics, University of Florida, Gainesville, Florida 32611-8440; Department of Chemistry and Biochemistry, University of California at San Diego, La Jolla, California 92093-0358

AUTHOR EMAIL ADDRESS (dhendrickson@ucsd.edu, and hill@phys.ufl.edu )

TITLE RUNNING HEAD: Single-Molecule Magnets: HFEPR of $Zn_3Ni$ Complex.

CORRESPONDING AUTHOR FOOTNOTE: Corresponding author: David N. Hendrickson[†] and Stephen Hill[‡]. [†]University of California at San Diego. [‡]University of Florida.

ABSTRACT: High-field electron paramagnetic resonance (HFEPR) spectra were collected at several frequencies for single crystal $[Zn_{3.91}Ni_{0.09}(hmp)_4(dmb)_4Cl_4]$ (**1**), where dmb is 3,3-dimethyl-1-butanol and hmp is the monoanion of 2-hydroxymethylpyridine. This crystal is isostructural to $[Ni_4(hmp)_4(dmb)_4Cl_4]$ (**2**), which has been characterized to be a single-molecule magnet (SMM) with fast quantum tunneling of its magnetization (QTM). The single $Ni^{II}$ ion zero-field-splitting (zfs) parameters $D_i$ [$= -5.30(5)$ cm$^{-1}$] and $E_i$ [$= \pm 1.20(2)$ cm$^{-1}$] in the doped complex **1** were evaluated by rotation of a crystal in three planes. The easy-axes of magnetization associated with the single-ion zfs interactions were also found to be tilted 15° away from the crystallographic $c$-direction. This inclination provides a possible explanation for the fast QTM observed for complex **2**. The single-ion zfs parameters are then related to the zfs parameters for the $Ni_4$ molecule by irreducible tensor methods to give $D = -0.69$ cm$^{-1}$ for the $S = 4$ ground state of the SMM, where the axial zfs interaction is given by $D\hat{S}_z^2$.

**Introduction.**

Single-molecule magnets (SMMs) are molecular nanomagnets that exhibit magnetization hysteresis below their blocking temperatures.[1-3] Several quantum effects have been observed in the magnetization dynamics of SMMs, such as tunneling of the direction of magnetization,[4,5] quantum phase interference,[6] and spin parity effects.[6,7] It has therefore been suggested that SMMs could some day be employed as the smallest magnetic memory unit[2,3] or in quantum computational devices.[8-10]

Polynuclear complexes of several first row transition metals have been reported to function as SMMs, manganese complexes being the most prevalent.[11] However, SMMs have also been identified with Fe,[12] V,[13] Cr,[14] and Ni.[15,16,17] In this paper, we focus on $Ni^{II}_4$ SMMs of the composition $[Ni_4(hmp)_4(ROH)_4Cl_4] \cdot S$, where $hmp^-$ is the monoanion of 2-hydroxymethylpyridine, R is some substituent, and S is a solvate molecule. As we previously communicated,[16,18] these $Ni^{II}_4$ SMMs exhibit magnetization versus magnetic field hysteresis loops that indicate the presence of a relatively fast rate of magnetization tunneling (a small coercive field in the loop) and, in certain cases, the presence of an exchange bias.[11,16] Detailed measurements have been carried out in order to determine the origin of the fast magnetization tunneling; the results will be given in several papers, including this one.

The three requirements for a molecule to be a SMM are: (1) a relatively large spin $S$ for the ground state; (2) an appreciable negative magnetoanisotropy, i.e. a dominant zero-field splitting (zfs) term, $D\hat{S}_z^2$ ($D < 0$), in the spin Hamiltonian; and (3) not too large a value for the tunnel splitting of the ground state (caused by terms in the spin Hamiltonian that do not commute with $\hat{S}_z$). The large spin and negative magnetoanisotropy determine the magnitude of the thermodynamic barrier for classical thermal activation, leading to reversal of the direction of magnetization for a molecule. It is important to emphasize that, even if a particular complex has a large barrier for magnetization reversal, it may not function as a SMM if the complex has a very fast rate of quantum tunneling of the direction of its magnetization. The $Ni^{II}_4$ SMMs have a spin $S = 4$ ground state and a negative magnetoanisotropy.[16,18,19] The question is why do they exhibit a fast rate of magnetization tunneling? In this and following papers we will answer this question.[19,20]

The $D$ value for a SMM results largely from the single-ion $D_i$ values for each of the metal ions in the molecule. The inter-ion magnetic exchange interactions in a $Ni^{II}_4$ complex give rise to a ground state with an axial zfs interaction $D\hat{S}_z^2$ that results from a projection of the single-ion zfs interactions at each of the four $Ni^{II}$ ions. $D$ values for $Ni^{II}$ complexes have been found to be both positive and negative.[21] Thus, the first question arising concerns the sign of the axial zfs parameter for the individual $Ni^{II}$ ions, and whether a positive single-ion $D_i$ value could project a negative $D$-value for the $S = 4$ ground state of a $Ni_4$ SMM? To answer this, we have prepared single crystals of $[Zn_{3.91}Ni_{0.09}(hmp)_4(dmb)_4Cl_4]$ (**1**), where dmb is 3,3-dimethyl-1-butanol. High-frequency electron paramagnetic resonance (HFEPR) experiments have then been carried out on a single-crystal of this doped complex in order to determine the single-ion spin zfs parameters ($D_i$, $E_i$, $g$, etc.) at each of the four possible $Ni^{II}$ sites for a $[Zn_3Ni(hmp)_4(dmb)_4Cl_4]$ molecule doped into a $[Zn_4(hmp)_4(dmb)_4Cl_4]$ crystal. We note that a similar methodology has recently been employed by Pardi et al., for a one-dimensional $Ni^{II}$ Haldane chain system.[22] However, to the best of our knowledge, this is the first such study by HFEPR for a SMM. The single-crystal measurements further enable us to determine the orientations of the principal magnetic axes at each of the $Ni^{II}$ sites, and to show how the $D$ value of the $S = 4$ $Ni^{II}_4$ SMM results from the $D_i$ and $E_i$ values ($i = 1$ to 4) for the four exchange-coupled $Ni^{II}$ ions. Finally, we investigate how the single-ion spin Hamiltonian parameters affect the thermodynamic barrier for magnetization reversal as well as the tunnel splitting for the $Ni_4$ SMM.

**Experimental Section.**

**Synthetic Procedure.** All operations were carried out in air. All reagents were purchased from Aldrich.

$[Zn_{3.91}Ni_{0.09}(hmp)_4(dmb)_4Cl_4]$ (**1**) – 1.23 g (9 mmol) of $ZnCl_2$, 0.24 g (1 mmol) of $NiCl_2 \cdot 6H_2O$, 1.09 g (10 mmol) of hmpH (2-hydroxymethylpyridine) and 0.54 g NaOMe (10 mmol) were dissolved in 80 mL MeOH and refluxed for 30 minutes. The resulting solution was then filtered and left at room temperature for over a week. Pale green prism-shaped crystals were then isolated. 1.6 g of these pale green crystals were collected by repeating the above procedures several times. The crystals were then suspended in 30 g of 3,3-dimethyl-1-butanol (dmb), and 40 mL of methylene chloride was added to dissolve the solids. The resulting solution was filtered and kept in an ambient environment for slow evaporation. Pale green bipyramidal-shaped crystals formed after two weeks with a overall yield of 10%. Anal. Calcd for **1**, $C_{48}H_{80}Cl_4N_4O_8Zn_{3.91}Ni_{0.09}$: C, 46.34; H, 6.48; N, 4.50. Found: C, 46.96; H, 6.55; N, 4.15. Selected IR data (KBr, $cm^{-1}$): 3310(br, s), 2960 (s), 2860 (s), 1600 (s), 1570 (s), 1480 (s), 1440 (s), 1400 (s), 1370 (s), 1290 (s), 1250 (m), 1220 (m), 1160 (s), 1080 (s), 1050 (s), 997 (s), 974 (s), 920 (m), 889 (w), 870 (w), 839 (m), 818 (s), 754 (s), 731 (s), 702 (w), 644 (s), 496 (m), 463 (m).

**X-ray Structure Determination.** Diffraction data for a crystal of $[Zn_{3.91}Ni_{0.09}(hmp)_4(dmb)_4Cl_4]$ were collected at 100(2) K with a Bruker Smart Apex CCD diffractometer equipped with Mo K$\alpha$ radiation ($\lambda = 0.71073$ Å). SADABS absorption correction was applied. The structure was solved by direct methods and refined on $F^2$ by a full-matrix least-squares procedure (Shelxtl Version 6.10, Bruker AXS, Inc., 2000). All atoms except the hydrogen were refined anisotropically. Hydrogen atoms were placed in calculated idealized positions. In the refinement, all metal atoms in the structure were treated as the Zn atom. Crystallographic data and details of X-ray study are given in Table 1.

**High frequency EPR (HFEPR).** A good sized crystal [approx. dimensions: $1.5 \times 1.5$ mm$^2$ square base; 2 mm height] of complex **1** was selected to perform the single-crystal HFEPR measurements. The orientation of the crystal was confirmed by determining crystal face indices based on X-ray diffractometry prior to the HFEPR measurements. The HFEPR spectrometer is equipped for conducting high sensitivity angle-dependent cavity perturbation measurements over a broad frequency range (40-350 GHz) and in the presence of a strong magnetic field (up to 17 tesla). A millimeter-wave vector network analyzer (MVNA) acts as a continuously tunable microwave source and phase sensitive detector (8-350 GHz), enabling simultaneous measurements of the complex cavity parameters at a rapid repetition rate (~10 kHz). A pumped $^4$He cryostat system was utilized to control the temperature of the experiment. A detailed description of the spectrometer can be found in the literature.[23,24]

**Inductively Coupled Plasma-Optical Emission Spectrometry (ICPOES).** A Perkin-Elmer Optima 3000 DV ICP-OES system with axially viewed plasmas was employed. The operation process of this system is described elsewhere.[25] The wavelengths of 213.86 nm for Zn and 231.60 nm for Ni were used to identify these elements. A quantity of 0.0375 g of complex (**1**) was diluted 118.5 times in 1% HNO$_3$ solution, and this solution was further diluted 160 times. The highest dilution factors were 196 ppb of Ni and 8460 ppb of Zn. Therefore a molar ratio of Zn:Ni= 97.74: 2.26 was obtained.

**Results and Discussion.**

**Preparation of the Ni$^{II}$ Doped Zn$^{II}_4$ Complex.** The goal was to prepare a crystal that is isostructural with [Ni$_4$(hmp)$_4$(dmb)$_4$Cl$_4$] (complex **2**), but where the Ni$^{II}$ ions are doped into a diamagnetic lattice. Through separate experiments, it was known that [Zn$_4$(hmp)$_4$(dmb)$_4$Cl$_4$] (complex **3**) could be prepared, where the Zn$^{II}$ ions obviously do not have unpaired electrons. Consequently, we prepared faintly green [Zn$_{3.91}$Ni$_{0.09}$(hmp)$_4$(dmb)$_4$Cl$_4$] (complex **1**) crystals. A relatively precise value of the Zn/Ni ratio in the crystals was obtained by ICPOES spectra which gave Zn:Ni = 97.74%:2.26%. Provided that there is little difference in the heats of formation for either the Zn$_4$ or Ni$_4$ complexes, then the Ni$^{II}$ ions in [Zn$_{3.91}$Ni$_{0.09}$(hmp)$_4$(dmb)$_4$Cl$_4$] should be randomly distributed in the crystal. Based upon this assumption, it is a relatively straightforward exercise to compute the probabilities for the formation of the Zn$_4$, Zn$_3$Ni, Zn$_2$Ni$_2$, ZnNi$_3$ and Ni$_4$ complexes, as a function $x$, in the formula [Zn$_{4x}$Ni$_{4-4x}$(hmp)$_4$(dmb)$_4$Cl$_4$] (see Fig. S1 in the on-line supplementary material). When $x = 0.977$, it is found that the Zn$_4$ and Zn$_3$Ni species make up 91% and 8% of the total population. Thus, the doped crystal is comprised of some $S = 1$ [Zn$_3$Ni$_1$(hmp)$_4$(dmb)$_4$Cl$_4$] complexes doped randomly into a diamagnetic Zn$_4$ host crystal. Since the Ni$^{II}$ ions are ferromagnetically coupled with each other in the Zn$_2$Ni$_2$, ZnNi$_3$, and Ni$_4$ complexes, one would expect ground states of $S = 2$, 3, and 4, respectively, for these three possibilities. The absence of significant HFEPR peak intensity for these spin values confirms that the predominant paramagnetic species present in the crystal is the $Zn_3Ni$ complex.

**X-ray Structure of [Zn$_{3.91}$Ni$_{0.09}$(hmp)$_4$(dmb)$_4$Cl$_4$].** Like the analogous Ni$_4$ complex **2**, [Zn$_{3.91}$Ni$_{0.09}$(hmp)$_4$(dmb)$_4$Cl$_4$] (**1**) crystallizes in the tetragonal I4$_1$/a space group with four molecules in the unit cell without solvate molecules (see Table 1). Fig. 1 shows the ORTEP plot of complex **1**. If each metal atom is treated as a Zn atom, molecule **1** has S$_4$ site symmetry with the metal ions occupying the alternating corners of the distorted cubane core. In addition to the chelating hmp$^-$ ligands, the central M$_4$O$_4$ core (M = Ni or Zn) is bonded to four Cl$^-$ anions and four 3,3-dimethyl-1-butanol ligands. A comparison of bond lengths and angles between complexes **1** and **2** is given in Table 2. As expected, the metal-ligand atom bond distances for the Zn$_{3.91}$Ni$_{0.09}$ complex are longer by 0.02-0.11 Å than those for the Ni$_4$ complex. The Ni-Cl bond length is, however, 0.002 Å longer than the Zn-Cl distance. The bond angles characterizing the Ni$_4$ and Zn$_{3.91}$Ni$_{0.09}$ complexes are similar. It seems reasonable that, in the crystal for complex **1**, the few Zn$_3$Ni complexes should be randomly distributed amongst the Zn$_4$ complexes. Also, it is clear that any spin Hamiltonian parameters characterized for the Ni$^{II}$ ions in complex **1** would be quite close in value to those found for the Ni$^{II}$ single ions in the Ni$_4$ crystal.

Crystals of **1** and **2** have the same space group and the same morphology, i.e. that of a rectangular-based bipyramid. The relationship between the unit cell directions and the crystal faces can be found in the on-line supplementary information (Fig. S2). The *c*-axis coincides with the bottom-to-top diagonal direction of the crystal, whereas the *a*- and *b*-axes are aligned with the edges of the rectangular base. The magnetic easy axis for complex **2** (the Ni$_4$ SMM) is found to be along the crystallographic *c*-axis.

**HFEPR Spectroscopy of [Zn$_{3.91}$Ni$_{0.09}$(hmp)$_4$(dmb)$_4$Cl$_4$] (1).** There are several reasons why we selected the doped crystal of complex **1** for single-crystal HFEPR experiments. First, the orientation of the crystal lattice relative to the faces of a crystal can easily be deduced. Second, the t-butyl groups on the dmb ligand provide considerable insulation between molecules in this complex. Therefore, intermolecular magnetic exchange interactions are minimized, as substantiated by the absence of any measurable exchange bias in the magnetization versus magnetic field hysteresis loop for the SMM Ni$^{II}$$_4$ complex (**2**).[16,18] Third, there are no solvated molecules in the crystals of complexes **1** and **2**. Disorder associated with any solvate molecules can give broad HFEPR signals as was seen for the methanol and ethanol analogues of complex **2**.[18,19] Thus, the crystals of complexes **1** and **2** give relatively sharp HFEPR signals, as will be discussed in detail in a later paper.[19]

Before the HFEPR spectra are discussed, it is worthwhile checking how many different metal sites are available for the individual Ni$^{II}$ ions. A stereoview of the molecular packing in the unit cell of complex **1** is shown in Fig. 2. At a first glance, it looks like there are two kinds of molecular orientations which are related by a four fold rotation axis. Thus, for each kind of cubane, with four Ni$^{II}$ sites available, there appear to be eight different single-ion orientations that need to be accounted for.



However, after a closer inspection, it is found that these two molecular orientations are actually related by an inversion operation. Therefore, because of the intrinsic inversion symmetry associated with the operators $D\hat{O}_2^0$ and $E\hat{O}_2^2$ in the spin-Hamiltonian, only four different orientations of the Ni$^{II}$ ions need to be considered.

The most direct way to deduce the orientations of the principal magnetic axes associated with the individual Ni$^{II}$ zfs interactions involves measuring the angle dependence of EPR spectra for several different planes of rotation. As will be seen below, such measurements were performed for three carefully chosen rotation planes. However, fits to such data contain many adjustable parameters, e.g. the various Ni$^{II}$ spin-Hamiltonian parameters ($D_i$, $E_i$, $g_i$, etc.), the Euler angles which relate the local magnetic axes to the crystallographic axes, and additional Euler angles which take into account minor mis-alignments between the actual and assumed rotation frames. For this reason, we first set out to estimate the Ni$^{II}$ spin-Hamiltonian parameters via frequency dependent measurements. Spectra were thus obtained between 58 GHz and 352 GHz with the DC field approximately aligned along the $c$-axis of the crystal, as displayed in Fig. 3; this orientation was deduced from preliminary angle-dependent data. Due to the four-fold symmetry of the crystal structure, one may assume that the angles ($\theta_i$) between the magnetic easy-axes for the four Ni$^{II}$ ion sites and the applied magnetic field are more-or-less the same for this field orientation. Thus, the frequency dependence of the spectra should roughly superimpose, with any splittings likely due to small deviations from perfect field alignment along the $c$-axis. Indeed, three distinct EPR absorptions are seen in the experimental spectra (see Fig. S3 in the on-line supplementary material). Each EPR absorption consists of several closely spaced peaks, which we believe are mainly due to a slight mis-alignment of the applied field. However, it is also likely that these splittings are caused by different micro-environments, as has been found for the parent [Ni(hmp)(dmb)Cl]$_4$ system.[18]

Fig. 3 displays the frequency dependence of the positions (in field) of the three EPR absorptions. Each data point is located at the center of an absorption, and the error bars indicate the range over which the fine-structure peaks are observed. Superimposed on the data are several simulations of the frequency dependence, obtained by taking difference frequencies between the $2S+1$ energy levels for $S=1$, as shown in Fig. 4. These triplet levels are labeled $T_x$, $T_y$ and $T_z$ in zero-field, in order of ascending energy; $T_z$ corresponds to the $M_S = 0$ state, while $T_x$ and $T_y$ correspond to the symmetric and antisymmetric combinations of the pure $M_S = \pm1$ states. At high-fields, the triplet states are labeled $T_{-1}$, $T_0$ and $T_{+1}$ in order of ascending energy; here, the subscripts refer to the spin projection onto the magnetic field quantization axis. Considering all possible transitions between these three levels, one expects three resonance branches ($A$, $B$ and $C$) which extrapolate to three different zero-field offsets (splittings). These offsets are sufficient for estimating $D_i$ and $E_i$: $D_i$ is responsible for the splitting between the upper $M_S = 0$ state and the $M_S = \pm1$ doublet; $E_i$ is responsible for the splitting of the $M_S = \pm1$ states. We note



that, to within the experimental uncertainty, single values for $D_i$ and $E_i$ are sufficient for fitting all of the data, as one would expect from the four-fold symmetry of the molecule.

Estimation of the *g*-factor requires knowledge of the field orientation relative to the magnetic axes at the individual $Ni^{II}$ sites. The dashed curves in Figs. 3 and 4 have been generated with the same $D_i$, $E_i$ and *g* values as the solid curves, but assumes that the single-ion easy axes are aligned with the applied magnetic field. In this situation, one sees that the $T_y$ and $T_z$ states pass right through each other, and that resonance *A* and *B* also pass right through each other. On the other hand, the data clearly indicate a level repulsion, which depends very sensitively on the field orientation. Thus, armed with approximate $D_i$ and $E_i$ values, one can adjust the field orientation (equivalent to tilting the single-ion easy axes away from the crystallographic *c*-axis) and the *z*-component of the *g*-factor until the behavior of the *A* and *B* resonances is accurately reproduced. This procedure provides a good first estimate of the tilting of the single-ion easy axes away from the crystallographic *c*-axis, i.e. $\approx 15°$.

The curves in Figs. 3 and 4 were actually generated using the final parameter set determined from the full angle dependence (see below and inset to Fig. 4). The two sets of solid curves in Fig. 3 illustrate the small effect of the transverse field component for this field orientation. Since we do not yet know the orientations of the hard and medium directions associated with each of the $Ni^{II}$ ions, we cannot know the orientation of the projection of the field within their hard-planes caused by the 15° tilting of the easy-axes. Thus, the two sets of curves illustrate the extreme cases corresponding to projection onto the hard and medium directions. One clearly sees that the lack of this information at this stage in the analysis does not dramatically influence our initial estimates of $D_i$, $E_i$, *g*, or the easy-axis tilting angle.

Finally, we comment on the selection rules governing the three branches of resonances in Fig. 3. First of all, as the sample is rotated in the horizontal field magnet, the relative orientation between the DC and AC magnetic fields ($H_0$ and $H_1$ respectively) changes. While it is possible to imagine a geometry in which $H_0$ and $H_1$ remain orthogonal, one cannot hope to achieve such a situation when working on high order modes of the cavity (*f* > 90 GHz). In addition, although the DC field is very homogeneous, the sample used in this study was rather large (approx. dimensions: 1.5×1.5 mm$^2$ square base; 2 mm height). Consequently, the $H_1$ field was not particularly uniform over the volume of the sample (note that $\lambda/2 \approx 0.5$ mm at 350 GHz), i.e. its orientation varies from one part of the sample to another. The relative orientations of $H_0$ and $H_1$ will also likely vary from one frequency to the next. Therefore, in general, one expects components of $H_1$ which are both parallel and perpendicular to $H_0$. Consequently, both $\Delta M_S = \pm 1$ and $\Delta M_S = \pm 2$ double quantum transitions should be expected. Indeed, the relative intensities of the different resonances differ noticeably for different frequencies (see Fig. S3 in the on-line supplementary material). In addition to this, there is strong mixing between the $T_y$ and $T_z$



states in the 3 T to 7 T range. This mixing also affects EPR selection rules, thereby allowing transitions which might otherwise be forbidden for a much higher symmetry geometry. It is for this reason that we simply label the three resonance branches *A*, *B* and *C*, i.e. it is meaningless to try to label them in terms of simple quantum numbers. Furthermore, we expect to observe all three resonances for essentially all field orientations and frequencies, i.e. no clear selection rules apply.

**Single-Ion Zero-Field Splitting.** In order to obtain more precise information concerning the orientations of the magnetic axes associated with the individual $Ni^{II}$ ions, further angle-dependent experiments were performed. Since the orientations of the crystallographic axes can be easily identified from the shape of the crystal, the sample can be manually aligned within the cavity for rotation in a desired crystallographic plane. However, such an alignment is never perfect, and it is thus necessary to introduce additional parameters to correct for small misalignments. In general, three parameters (Euler angles: $\alpha$, $\beta$, $\gamma$) are required in order to relate any two coordinate systems; these angles are defined in Fig. S4 in the on-line supplementary material and in ref. [26]. We correct for any mis-alignment of the sample within the cavity by defining a set of Euler angles $C_\alpha$, $C_\beta$ and $C_\gamma$, which relate the crystallographic (*c*) and laboratory (*l*) coordinate frames, i.e. $(a, b, c) \leftrightarrow (X, Y, Z)$.

We can now approximate the $[Zn_3Ni(hmp)_4(dmb)_4Cl_4]$ molecule as a cube with its edges aligned with the crystallographic *a*, *b*, and *c* axes, as shown in Fig. 5, with four possible positions (labeled 1 to 4) for the lone $Ni^{II}$ ion. We note that the actual molecule has a distorted cube structure. Next, we must relate each of the four single-ion coordinate frames $(x_i, y_i, z_i)$ with the crystallographic axes $(a, b, c)$; the subscripts *i* refer to the four positions. Due to the four-fold symmetry of the crystal, only a single set of Euler angles $(S_\alpha, S_\beta, S_\gamma)$ is needed to completely specify these four coordinate systems, i.e. $S_\alpha$, $S_\beta$ and $S_\gamma$ specify the single-ion axes at one of the sites (e.g. $i = 1$), and the remaining three are related by the crystallographic symmetry group. The single-ion axes illustrated in Fig. 5 have been oriented approximately as deduced from these studies. For the ensuing discussion, we use the angles $\theta$ and $\phi$ to denote polar angles; we use the subscripts *l* and *c* to refer to the laboratory and crystal coordinate systems, respectively, and subscripts $i$ (= 1 to 4) to denote the four single-ion coordinate systems (Fig. 5).

Angle-dependent measurements were performed for three separate planes of rotation; the blueprints for the three experiments are illustrated in Fig. S5 in the on-line supplementary material. For orientation 1, we rotated the field approximately in the crystallographic *ac*-plane [or the *bc*-plane, since it is indistinguishable by symmetry], i.e. in a plane parallel to one of the vertical cube faces in Fig. 5. For orientation 2, we rotated the field approximately in the crystallographic *ab*-plane, i.e. in a plane parallel to the horizontal cube faces in Fig. 5. Finally, for orientation 3, we rotated the field in a plane



approximately intersecting the crystallographic *ac*- and *bc*-planes, i.e. a vertical plane crossing the horizontal diagonals of the cube in Fig. 5. These geometries are also depicted as insets to Fig. 6. A frequency close to 120 GHz was carefully selected for these experiments. This sits just at the lower edge of a gap between the *B* and *C* resonance branches (Fig. 3). In fact, 120 GHz is right at the zfs between the $T_y$ and $T_z$ levels (Fig. 4). Consequently, resonance *C* is always observed as a broad absorption at zero-field and resonance *B* is always found t higher frequencies. Meanwhile, the position of resonance *A* depends strongly on the field orientation at 120 GHz, moving from 1.5 T upwards to the maximum available field of 7 T for field orientations close to the hard magnetic directions. In this way, 120 GHz provides the lowest frequency (hence the best signal-to-noise) with which we can focus on the angle-dependence of a single resonance, i.e. resonances *B* and *C* do not interfere with the angle-dependence of resonance *A* over the full $4\pi$ radians.

Fig. 6 displays the angle dependence of peaks corresponding to resonance *A* (at 2.5 K and $f \approx 120$ GHz) for the three chosen planes of rotation. Representative experimental spectra are displayed in Fig. S6 in the on-line supplementary material, and the precise frequencies are listed in the figure caption. For each plane of rotation, four resonance branches are observed corresponding to the four Ni$^{II}$ sites, thus confirming the earlier conclusion concerning the non-colinearity of the magnetic axes at each site. The solid curves are fits to the data; we describe the fitting procedure in detail below. In Figs. 6(a) and (c), the experimental angle ('offset') has been calibrated so that the high-symmetry points occur at $\theta_l = 0°$ and $\theta_l = 90°$. For both of these orientations, the Euler angle $C_\beta \approx 0$. Consequently, the angles $\theta_l$ and $\theta_c$ are approximately equivalent. Meanwhile, the angle in Fig. 6(b) (orientation 2) is referenced to the approximate orientation of the crystallographic *a*- (or *b*-) axis, as deduced via microscope images of the sample mounted in the cavity. We estimate the error of this calibration to be $\pm 5°$.

Before discussing the data, we briefly review the Hamiltonian for spin $S = 1$. We can write the Hamiltonian operator in matrix form as follows:

$$\hat{H}_i = \begin{pmatrix} D + \mu_B g_z B \cos(\theta_i) & A_i & E \\ A_i^* & 0 & A_i \\ E & A_i^* & D - \mu_B g_z B \cos(\theta_i) \end{pmatrix},$$

where (1)

$$A_i = 0.707 \mu_B B \left[ g_x \sin(\theta_i) \cos(\phi_i) - i g_x \sin(\theta_i) \sin(\phi_i) \right].$$



It should be noted that the angles $\theta_i$ and $\phi_i$ involved in this Hamiltonian represent the spherical polar coordinates referenced to each of the individual Ni$^{II}$ spin coordinate frames ($x_i$, $y_i$, $z_i$), as depicted in Fig. 5. All other terms in Eq. (1) represent zfs interactions, which we find to be the same for all four ions. One must then independently transform from the lab coordinates into each spin coordinate system using the two Euler transformations described earlier. For each experimental orientation, this involves a separate set of $C$ parameters. However, our fitting procedure is designed to obtain a single set of $S$ parameters.

The next step involves solving for the eigenvalues (frequencies) for all four spin sites, as a function of the applied magnetic field strength and its orientation. We note that the field was only ever rotated about a single axis for a particular experiment; thus, the field strength ($B$), and its orientation ($\theta_l$) in the lab frame, represent the only variables in this procedure. Upon doing so, we obtain four 3D surfaces (frequency, field, orientation) – one for each Ni$^{II}$ site. For rotations in high-symmetry crystallographic planes, the number of surfaces may be reduced due to degeneracies. However, we shall see that, in general, the experimental geometry was such that these degeneracies are completely lifted. Fig. 7 shows a pair of doubly degenerate surfaces for rotation exactly in the *ab*-plane, i.e. very close to the rotation plane for orientation 2 [Fig. 6(b)]. We then perform a 'virtual' experiment, where we pick a measurement frequency ($f$) and solve for the intersection between the constant $f$ surface and the four 3D surfaces. The resulting curves describe the angle dependence of the spectra for a particular set of $C$, $S$ and zfs parameters ($D_i$, $E_i$ and $g$). The final, and most complicated task involves performing a least squares fit to the full data set (orientations 1 to 3).

As already noted, the four resonance branches seen in Fig. 6 originate from each of the four Ni$^{II}$ sites (Fig. 5). All degeneracies were avoided because of small mis-alignments of the sample within the cavity for each rotation plane, i.e. finite $C$ parameters (Euler angles). However, it is apparent that this alignment was fairly good in the case of orientation 1, because the two high-field branches are almost degenerate. In zero-magnetic field, the magnetic dipole transitions for all four Ni$^{II}$ ions must be degenerate due to the symmetry equivalence of the four sites. Application of a magnetic field breaks this symmetry. However, a magnetic field along the *c*-axis will not lift any degeneracies due to the axial symmetry, while a field applied within the *ab*-plane will only partially lift the four-fold degeneracy, i.e. in general, the spectrum will split into two branches, as illustrated in Fig. 7. Note that the curves for orientations 1 and 3 [Figs. 6(a) and (c)] intersect at a single point at $\theta_l = 0°$ ($\approx \theta_c$), and at two points for $\theta_l = 90°$. The two branches arise because of the *E*-term: when the field is roughly parallel to the $x_i$-axes of two of the ions, it must be approximately orthogonal to the $x_i$-axes of the other two. Thus, it is the combination of the Zeeman and *E*-terms that lift the degeneracy.



As soon as the field is tilted away from either the *c*-axis, or the *ab*-plane, the degeneracy will in general be completely lifted. This is why four branches are seen for each experiment. In particular, for orientation 2, the rotation plane was inclined at an angle $C_\beta = 9.8°$ away from the *ab*-plane, causing the splitting of the two main branches. The dashed curves in Fig. 6(b) reveal the average behavior of each split branch. The minima of the dashed curves, which are separated by exactly 90° (as required by symmetry), correspond to the directions of the projections of the medium axes of the individual $Ni^{II}$ ions ($y_i$) onto the *ab*-plane. Meanwhile, the projections of the hard axes ($x_i$) correspond to the maxima in the dashed curves [not shown in Fig. 6(b)] which, to within the experimental resolution, occur at the same angles. Thus, from Fig. 6(b), we can conclude that the hard and medium axes of the $Ni^{II}$ ions ($x_i$ and $y_i$) project onto the *ab*-plane along directions given by $\phi_c = 70\pm5° + (n \times 90°)$ [$n$ = integer] relative to the *a*-axis; the ±5° results from the uncertainty in the alignment of the crystal within the cavity.

Based on the near degeneracy of the two high-field branches seen for orientation 1 [Fig. 6(a)], one can conclude that the single-ion easy-axes tilt approximately along the crystallographic *a* and *b*-directions (±5°), i.e. $S_\alpha \approx 0°$, 90°, etc. It is then straightforward to deduce the single-ion easy-axis tilting angle ($S_\beta$) from the separation of the low-field (~3.75 T) maxima in Fig. 6(a). This maximum separation of 30°±1° corresponds to the angle between hard planes associated with $Ni^{II}$ sites which are related by a 180° rotation (see Fig. 5), implying that the easy-axes for each $Ni^{II}$ ion are tilted 15±0.5° (= $S_\beta$) away from the *c*-axis, as previously inferred on the basis of the frequency dependence. Finally, we can conclude from the low-field maxima in Fig. 6(a) that the magnetic field intersects the hard planes associated with the two $Ni^{II}$ ions responsible for these maxima closer to their medium directions (~20° away) than the hard directions. Thus, based on the discussion of Fig. 6(b) above, we infer that the Euler angle $S_\gamma = 70\pm5°$ (or 250±5°) for the ions with $S_\alpha \approx 0°$ (or 180°), and $S_\gamma = 160\pm5°$ (or −20±5°) for the ions with $S_\alpha \approx 90°$ (or 270°). For more detailed explanation, refer to the on-line supplementary material.

The optimum single-ion zfs parameters obtained from these angle-dependent measurements, as well as frequency dependent studies performed with the magnetic field applied along the crystallographic *c*-axis (Fig. 3), and the *a*/*b*-axes (not shown), yield: $D_i = -5.30(5)$ cm$^{-1}$, $E_i = \pm1.20(2)$ cm$^{-1}$, $g_z = 2.30(5)$, and $g_x = g_y = 2.20(5)$. To within our experimental accuracy, the obtained *g*-anisotropy is marginal. Based on the obtained $D_i$ and $E_i$ values, one would expect a difference between $g_z$ and $g_{xy}$ [=½($g_x + g_y$)] of order of 0.02, and a difference between $g_x$ and $g_y$ on the order of 0.01.[22] This is *not* inconsistent with the obtained values. Unfortunately, the relatively broad EPR signals do not permit a more precise determination of the *g*-anisotropy.

**Temperature dependence.** We conclude this section by examining the temperature dependence of the *A* resonance. Fig. 8 displays data obtained at temperatures between 2.5 K and 10 K, at a frequency of 119.9 GHz; the field is oriented approximately along the *c*-axis of the crystal. The inset to Fig. 8



shows the energy level diagram for $S = 1$, with a negative uniaxial $D_i$ [$= -5.30(5)$ cm$^{-1}$] parameter [also $E_i = \pm 1.20(2)$ cm$^{-1}$]. As can be seen from the figure, the $A$ transition occurs from the ground state. Thus, one should expect the intensity of this transition to increase, as the temperature is decreased. This is precisely the behavior seen in the main part of Fig. 8. In contrast, if the sign of $D_i$ was positive, the energy level diagram would be inverted. In this case, the $A$ transition would occur from an excited state, and its intensity should decrease at the lowest temperatures. Since such a behavior is not observed, we can conclude from the temperature dependent studies that the sign of $D_i$ must be negative.

**Zero-Field Splitting of the $S = 4$ Ground State of the Ni$_4$ SMM.** The SMM complex [Ni$_4$(hmp)$_4$(dmb)$_4$Cl$_4$] has a spin $S = 4$ ground state as a result of the ferromagnetic coupling of the four Ni$^{II}$ ions. The barrier for magnetization reversal is approximately equal to $S^2|D|$, where $D$ gauges the axial zero-field splitting, $D\hat{S}_z^2$, in the $S = 4$ ground state. In this section it is shown how the single-ion zfs interaction for the single Ni$^{II}$ ion in the [Zn$_3$Ni(hmp)$_4$(dmb)$_4$Cl$_4$] molecule doped into the lattice of diamagnetic Zn$^{II}_4$ molecules gives rise to the axial zfs interaction for the $S = 4$ ground state of the Ni$^{II}_4$ SMM.

In the previous section, the zfs parameters for the Ni$^{II}$ ion in [Zn$_3$Ni(hmp)$_4$(dmb)$_4$Cl$_4$] were evaluated by HFEPR to be $D_i = -5.30(5)$ cm$^{-1}$ and $E_i = \pm 1.20(2)$ cm$^{-1}$. The zfs matrix in each local Ni$^{II}$ coordinate system is of the form:

$$\overline{\overline{D}}_{local} = \begin{bmatrix} -D_i/3 + E_i & 0 & 0 \\ 0 & -D_i/3 - E_i & 0 \\ 0 & 0 & 2D_i/3 \end{bmatrix} \quad (2)$$

Hereafter, the double bar symbol above each capital letter represents a 3×3 matrix. This matrix has to be converted from the local coordinate framework to the crystal lattice coordinate framework. This transformation is made by means of the Euler matrix $\overline{\overline{A}}$, where the Euler angles $(S_\alpha, S_\beta, S_\gamma) = (0°, 15°, 70°)$, and is given as:

$$\overline{\overline{D}}_{crystal} = \overline{\overline{A}}^{-1} \overline{\overline{D}}_{local} \overline{\overline{A}}. \quad (3)$$

In the Ni$_4$ complex there are four Ni$^{II}$ ions and, if we define $\overline{\overline{D}}_1 \equiv \overline{\overline{D}}_{crystal}$ for Ni$^{II}$ ion number 1, then the total zfs matrix for a Ni$^{II}_4$ molecule is given by eq(4).

$$\overline{\overline{D}}_{total} = d_1\overline{\overline{D}}_1 + d_2\overline{\overline{D}}_2 + d_3\overline{\overline{D}}_3 + d_4\overline{\overline{D}}_4 + d_{12}\overline{\overline{D}}_{12} + d_{13}\overline{\overline{D}}_{13} + d_{14}\overline{\overline{D}}_{14} + d_{23}\overline{\overline{D}}_{23} + d_{24}\overline{\overline{D}}_{24} + d_{34}\overline{\overline{D}}_{34}. \quad (4)$$



In this equation, the $\overline{\overline{D}}_i$ ($i = 1, 2, 3, 4$) are the zfs matrices for the individual ions, $\overline{\overline{D}}_{ij}$ ($i, j = 1, 2, 3, 4$) are the asymmetric dipolar couplings between metal centers $i$ and $j$, and the $d_i$ and $d_{ij}$ are their expansion coefficients. The values of $d_i$ and $d_{ij}$ are calculated by irreducible tensor methods.[27-29] In the $Ni^{II}_4$ cubane molecule, $S_1 = S_2 = S_3 = S_4 = 1$, and $S = 4$. Thus,

$$\overline{\overline{D}}_{total} = \tfrac{1}{28}\overline{\overline{D}}_1 + \tfrac{1}{28}\overline{\overline{D}}_2 + \tfrac{1}{28}\overline{\overline{D}}_3 + \tfrac{1}{28}\overline{\overline{D}}_4 + \tfrac{2}{28}\overline{\overline{D}}_{12} + \tfrac{2}{28}\overline{\overline{D}}_{13} + \tfrac{2}{28}\overline{\overline{D}}_{14} + \tfrac{2}{28}\overline{\overline{D}}_{23} + \tfrac{2}{28}\overline{\overline{D}}_{24} + \tfrac{2}{28}\overline{\overline{D}}_{34} \qquad (5)$$

Since the $\overline{\overline{D}}_{ij}$ are mainly from dipole-dipole interactions, for simplicity we ignore their contribution to $\overline{\overline{D}}_{total}$. This is reasonable, since the $\overline{\overline{D}}_{ij}$ are on the order of 0.1 cm$^{-1}$, whereas in the $Ni^{II}_4$ complex **3**, $\overline{\overline{D}}_1$ is about 5 cm$^{-1}$. This gives eq.(6):

$$\overline{\overline{D}}_{total} = \tfrac{1}{28}\overline{\overline{D}}_1 + \tfrac{1}{28}\overline{\overline{D}}_2 + \tfrac{1}{28}\overline{\overline{D}}_3 + \tfrac{1}{28}\overline{\overline{D}}_4 \qquad (6)$$

The $\overline{\overline{D}}_i$ matrices for the other three $Ni^{II}$ ions numbered 2, 3, and 4 can be related to $\overline{\overline{D}}_1$ for $Ni^{II}$ ion number 1 by employing the matrix for the $S_4$ symmetry operation. The transformations for $\overline{\overline{D}}_2$, $\overline{\overline{D}}_3$, and $\overline{\overline{D}}_4$ are then given by eq.(7):

$$\overline{\overline{D}}_2 = \overline{\overline{S}}_4^{\,1}\,\overline{\overline{D}}_1\,\overline{\overline{S}}_4^{\,-1}$$

$$\overline{\overline{D}}_3 = \overline{\overline{S}}_4^{\,2}\,\overline{\overline{D}}_1\,\overline{\overline{S}}_4^{\,-2} \qquad (7)$$

$$\overline{\overline{D}}_4 = \overline{\overline{S}}_4^{\,3}\,\overline{\overline{D}}_1\,\overline{\overline{S}}_4^{\,-3}$$

The numerical results for $\overline{\overline{D}}_{total}$, with the positive single-ion $E_i$ value, are given as follows:

$$\overline{\overline{D}}_{total} = \begin{bmatrix} 6.8998 & 0 & 0 \\ 0 & 6.8998 & 0 \\ 0 & 0 & -13.7995 \end{bmatrix} \qquad (8)$$

It is satisfying that $\overline{\overline{D}}_{total}$ is a traceless diagonal matrix with (1,1) = (2,2). From Eq. (2), this implies that $E = 0$ for the $S = 4$ ground state of the $Ni^{II}_4$ SMM, as expected for this complex, which has $S_4$ site symmetry, requiring that $E = 0$. From eqs. (2) and (8) we can calculate that $D = -0.69$ cm$^{-1}$ for the positive single-ion $E_i$ value (and $D = -0.66$ cm$^{-1}$ for a negative single-ion $E_i$ value) for the $S = 4$ ground state. Given the approximations made in this calculation, agreement with the experimentally determined $D$ value of $-0.600$ cm$^{-1}$ is remarkably good.

**Origin and Importance of ZFS for the $S = 4$ Ground State of the $Ni^{II}_4$ SMM.** It was shown in the last section how the single-ion zfs interactions at each metal ion site contribute to the zfs interaction for the ground state of a SMM. For metal ions such as $Ni^{II}$, that have an orbitally non-degenerate ground state in a low-symmetry coordination site, the origin of the zfs interaction is the admixture of excited states into the ground state wave function that result from spin-orbit coupling ($\lambda \hat{L} \cdot \hat{S}$). These single-ion zfs effects are parameterized in a spin Hamiltonian with terms such as $D_i \hat{S}_z^2$ and $E_i(\hat{S}_x^2 - \hat{S}_y^2)$ – the axial and rhombic zfs interactions. Here, we retain the subscript $i$ to distinguish between single-ion and SMM zfs parameters. The magnitude and sign of the $D_i$ parameter (also the $E_i$ parameter) depend on the magnitude and sign of the spin-orbit coupling constant $\lambda$, and on how close in energy the excited states are that are spin-orbit admixed with the ground state. The single-ion $D_i$ value for a $Ni^{II}$ complex has been found to be positive in some complexes, but negative in others.[21,22] For six-coordinate $Ni^{II}$ complexes with six equivalent ligands, $D_i$ values have been reported in the range of –3.05 to +1.70 cm$^{-1}$. In a recent paper,[30] Krzystek et al. used HFEPR to determine $D_i = +13.20(5)$ cm$^{-1}$ for the four-coordinate tetrahedral $Ni^{II}$ complex $Ni(PPh_3)_2Cl_2$; the complex $Ni(PPh_3)_2Br_2$ gave $D_i = +4.5(5)$ cm$^{-1}$. Thus, four-coordinate complexes generally have larger $|D_i|$ values than do six-coordinate complexes, because the excited states are closer in energy to the ground state in the case of the former. Consequently, the magnitude and sign of $D_i$ for a $Ni^{II}$ complex reflects the number and geometry of ligands. In addition, the orientations of the principal magnetic axes associated with the single-ion zfs interaction reflect the positioning of the ligand atoms. In other words, the $z_i$ direction associated with the $D_i \hat{S}_z^2$ axial zfs interaction might correspond to a metal-ligand bond direction, or some other orientation depending on the positioning and symmetry of the ligands. The magnitude, sign and direction of the single-ion zfs interaction ultimately determines the zfs interaction in a $Ni^{II}_4$ SMM. If the $D_i$ value for a $Ni^{II}$ ion is negative, then it is necessary that the magnetic $z_i$-axis of that ion not be inclined too far away from the magnetic $z$-axis of the $Ni^{II}_4$ SMM ($c$-axis in this case). Alternatively, the single ion $D_i$-value could be positive. However, in this situation, the single-ion magnetic $z_i$-axis should be oriented close to the hard plane ($xy$-plane) in order for the positive $D_i$ value to project a negative contribution to the $D$ value for the $Ni^{II}_4$ SMM.

There are two important ways in which the single-ion zfs interactions (both $D_i$ and $E_i$) affect the low-temperature magnetization dynamics of a SMM. In order for a SMM to have a potential energy barrier for magnetization reversal, the SMM must have a negative $D$ value. The $Ni^{II}_4$ SMMs have spin $S = 4$ and the barrier height $U_{eff}$ is $S^2|D| \approx 10$ cm$^{-1}$. One way for the $Ni^{II}_4$ SMM to reverse its magnetization direction from "spin-up" ($M_S = -4$) to "spin-down" ($M_S = 4$) is via classical thermal activation over this barrier. Thus, it is clear that the barrier height will affect the low-temperature ($k_BT < S^2|D|$) magnetization dynamics. This barrier height is ultimately related to the tensor projection of the single-ion zfs parameters $D_i$ and $E_i$.

The single-ion zfs interaction can play an even more important role in determining the quantum (as opposed to classical) magnetization dynamics of a $Ni^{II}_4$ SMM. At very low temperatures ($k_BT \ll S^2|D|$), where SMMs exhibit blocking of their magnetization, quantum tunneling may become the dominant mechanism of magnetization reversal. In this case, one observes a temperature-independent magnetization relaxation rate, involving tunneling between the $M_S = -4$ and $M_S = +4$ ground states. Transverse interactions in the spin Hamiltonian (terms that do not commute with $\hat{S}_z$) cause this tunneling at a rate which is governed by the tunnel-splitting interaction (raised to some power) which lifts the degeneracy between the $M_S = -4$ and $M_S = +4$ states. How effective this interaction is at causing tunneling depends very sensitively on the symmetry of the spin Hamiltonian and on the magnitude of the spin. Transverse interactions can arise from transverse internal or external magnetic fields (i.e. fields with components in the *xy*-plane of the SMM), or from zfs interactions which contain the operators $\hat{S}_x$ and $\hat{S}_y$. As we will show in a later paper, where HFEPR data are presented for a single-crystal of a $Ni^{II}_4$ SMM, the quartic zfs interaction $B_4^4 \hat{O}_4^4$ causes the rapid magnetization tunneling. It has been determined in this paper that the $Ni^{II}$ ion easy-axes are tilted 15° away from the crystallographic *c*-axis (i.e., the molecular easy-axis) for the $S = 4$ ground state of the $Ni_4$ SMM. A significant $E_i$ term is also found for the single $Ni^{II}$ ion. We believe that these factors, when combined with the four-fold symmetry, project out a large $B_4^4$ term for the $S = 4$ ground state. It is shown in the later paper exactly how this transverse fourth-order $B_4^4 \hat{O}_4^4$ term leads to fast ground-state quantum magnetization tunneling in the $Ni_4$ SMM. In summary, the single-ion zfs interactions in a $Ni^{II}_4$ SMM determine both the potential-energy barrier for magnetization reversal and, more importantly, the rate of magnetization tunneling.

ACKNOWLEDGMENT. This work was supported by National Science Foundation (CHE0095031, CHE0350615, DMR0103290, DMR0239481).

**Supporting Information Available.** X-ray crystallographic files in CIF format are available for complexes **1**. This material is made available free of charge via the Internet at http://pubs.acs.org..

# References

[†]University of California at San Diego. [‡]University of Florida.

**Fig. 1.** ORTEP plot of [Zn$_{3.91}$Ni$_{0.09}$(hmp)$_4$(dmb)$_4$Cl$_4$] at the 30% probability level. Intramolecular O-H⋯Cl H-bonds are shown by dashed lines.

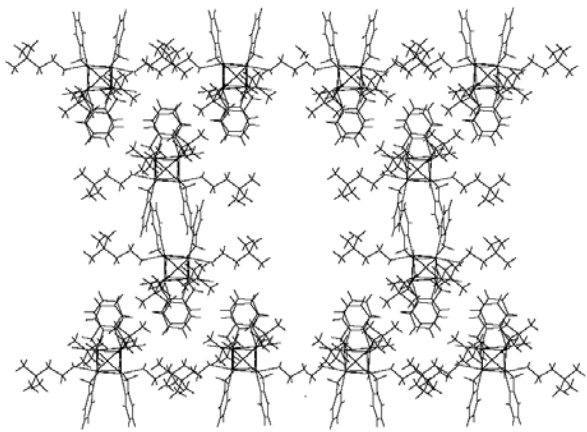

**Fig. 2.** Stereo plot of the molecule packing in [Zn$_{3.91}$Ni$_{0.09}$(hmp)$_4$(dmb)$_4$Cl$_4$].

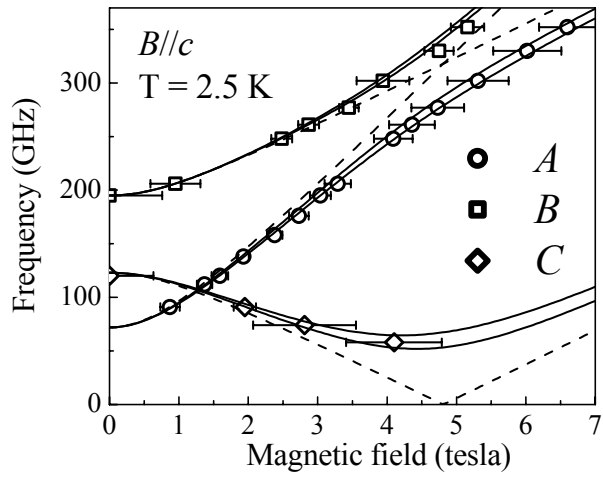

**Fig. 3.** Frequency dependence of the positions (in field) of the three resonance branches. Each point is located at the center of a resonance, and the error bars represent the range over which fine structures are observed. Superimposed on the data are several simulations of the frequency dependence of the resonance positions, obtained by taking difference frequencies between the $2S+1$ energy levels for $S = 1$, as shown in Fig. 4. See main text for explanation of the simulations, and Fig. 4 for labeling of the various transitions ($A$, $B$, $C$).

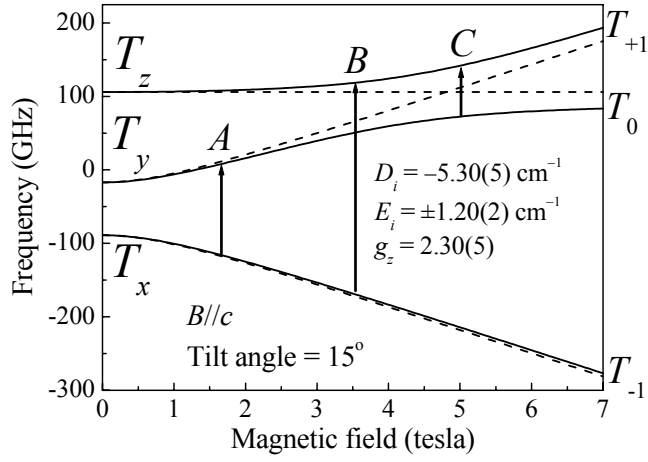

**Fig. 4.** Energy level diagram for $S = 1$, assuming a negative uniaxial $D_i$ parameter. The spin Hamiltonian parameters, corresponding to the obtained values for [NiZn$_3$(hmp)$_4$(dmb)$_4$Cl$_4$], are indicated in the figure. The solid curves correspond to the case where the magnetic field is applied parallel to the $c$-axis of the crystal [(001) direction]. The dashed curve corresponds to the case where the magnetic field is applied parallel to the local easy-axis of one of the Ni$^{II}$ ions. The levels have been labeled in zero-field according to the scheme described in the main text, and the transitions observed in Fig 3 are represented by the vertical arrows.

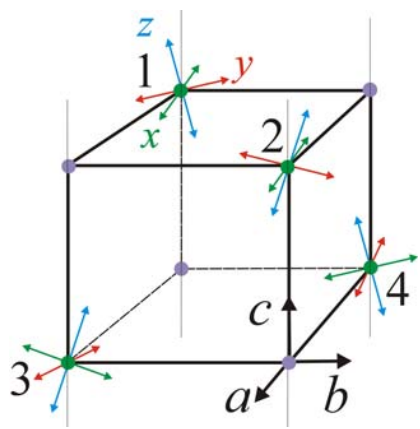

**Fig. 5.** A schematic representation of the four possible principle single-ion magnetic axes ($x_i$, $y_i$, $z_i$) tilted with respect to the crystallographic axes ($a$, $b$, $c$). A solid cube is drawn with each of the edge aligned with the crystal axes to guide the eyes and to show their relationship. We note that the real [NiZn$_3$(hmp)$_4$(dmb)$_4$Cl$_4$] molecule has a distorted cube structure.

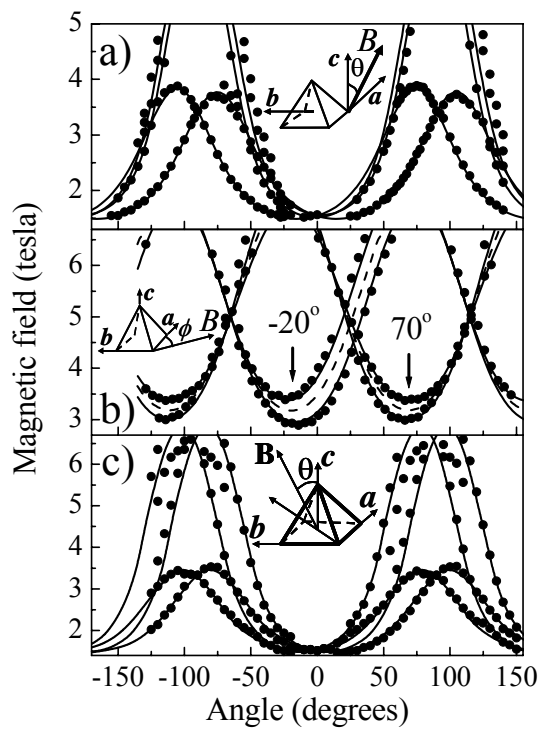

**Fig. 6.** Angle-dependence of the EPR peak positions corresponding to resonance *A* (Fig. 4) for: (a) orientation 1, field rotation approximately in the *ac*-plane (or *bc*-plane); (b) orientation 2, field rotation approximately in the *ab*-plane; and (c) orientation 3, field rotation approximately in the plane containing the (110) and (001) crystallographic directions. The inset depicts the experimental geometry. In each case, the temperature was 2.5 K, and the frequencies were (a) 119.9 GHz, (b) 122.8 GHz, and (c) 117.7 GHz. The experimental spectra can be observed in the on-line supplementary material. The solid curves are fits to the data, and the dashed curves in (b) represent the averages of the pairs of almost degenerate curves (see main text). The directions of the projections of the single-ion hard and medium directions onto the *ab*-plane have been indicated in (b).

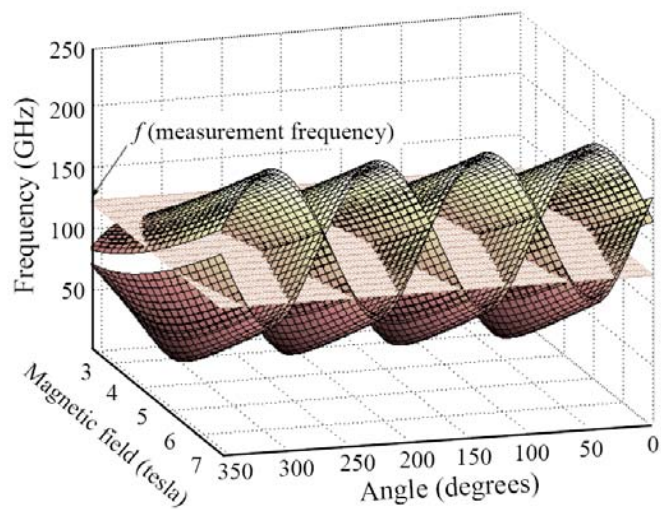

**Fig. 7.** Simulated resonance surfaces for field rotation exactly in the *ab*-plane. Two branches of resonances are seen (see main text for explanation). The angle dependence of the resonance positions is found via the intersection of the surfaces with a horizontal plane corresponding to the experimental measurement frequency. The simulations assume the parameters given in Fig. 4.

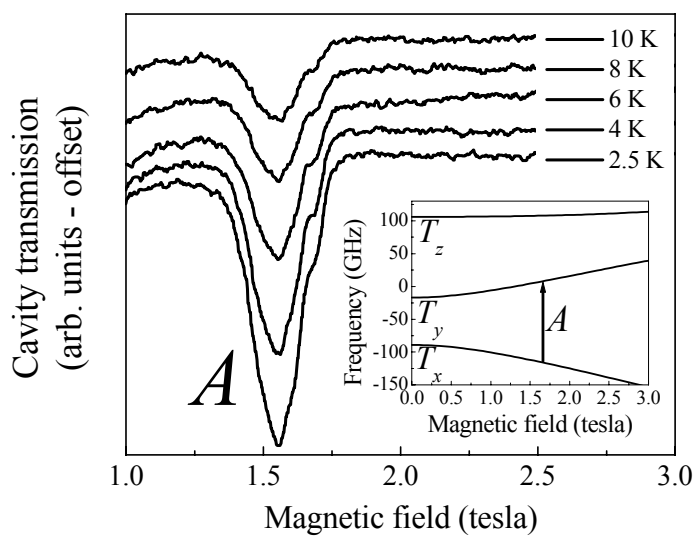

**Fig. 8.** Temperature dependence of the *A* resonance between 2.5 K and 10 K, and at a frequency of 119.9 GHz; the field is oriented approximately along the crystallographic *c*-axis. The inset shows the energy level diagram for $S = 1$, with a negative uniaxial $D$ [$= -0.530(5)$ cm$^{-1}$] parameter [also $E = \pm 1.20(2)$ cm$^{-1}$]. As can bee seen from the figure, the *A* transition occurs from the ground state.

**Table 1.** Crystallographic Data for [Zn$_{3.91}$Ni$_{0.09}$(hmp)$_4$(dmb)$_4$Cl$_4$] (complex **1**)

| | |
|---|---|
| formula[a] | C$_{48}$H$_{80}$Cl$_4$N$_4$O$_8$M |
| fw, g/mol | 1244.44 |
| temperature | 100(2) K |
| Space group | *I4$_1$/a* |
| *a*, Å | 12.9165(4) |
| *b*, Å | 12.9165(4) |
| *c*, Å | 35.038(2) |
| α= β= γ | 90° |
| Volume | 5845.7(4) Å$^3$ |
| Z, Z' | 4, 0.25 |
| F(000) | 2592 |
| Density Calculated | 1.414 g·cm$^{-3}$ |
| absorption coefficient | 1.854 mm$^{-1}$ |
| absorption correction | SADABS |
| transmission max./min. | 0.7684/0.6062 |
| Reflections, measured | 18318 |
| Reflections, independent | 3355 [Rint=0.0181] |
| data/restrains/parameters | 3355/0/234 |
| goodness-of-fit on F$^2$ | 1.055 |
| R indices [I>2σ(I)] | R1=0.0214, wR2= 0.0546 |
| R indices (all data) | R1=0.0233, wR2= 0.0554 |

[a.] M represents Zn or Ni.

**Table 2.** Comparison of selected bond length (Å) and bond angle (°) for complexes **1** and **2**. (where M represents Zn or Ni in **1**, and Ni in **2**)

|  | complex **2** | complex **1** |
| --- | --- | --- |
| M(1)-O(1)#1 | 2.0510(13) | 2.0709(9) |
| M(1)-O(1)#2 | 2.1018(13) | 2.1245(9) |
| M(1)-O(1) | 2.0374(13) | 2.1338(9) |
| M(1)-O(2) | 2.0893(14) | 2.1972(10) |
| M(1)-N(1) | 2.0648(16) | 2.1246(11) |
| M(1)-Cl(1) | 2.3593(5) | 2.3377(3) |
| O(1)#1-M(1)-O(1)#2 | 80.73(5) | 81.02(4) |
| O(1)#1-M(1)-N(1) | 161.13(6) | 156.35(4) |
| O(1)#2-M(1)-N(1) | 101.79(5) | 103.41(4) |
| O(1)#1-M(1)-O(1) | 82.29(4) | 80.80(4) |
| O(1)#2-M(1)-O(1) | 80.10(5) | 79.92(4) |
| N(1)-M(1)-O(1) | 79.78(6) | 77.22(4) |
| O(1)#1-M(1)-O(2) | 85.13(5) | 83.07(4) |
| O(1)#2-M(1)-O(2) | 165.09(5) | 162.74(4) |
| N(1)-M(1)-O(2) | 90.09(6) | 88.63(4) |
| O(1)-M(1)-O(2) | 93.38(5) | 90.97(4) |
| O(1)#1-M(1)-Cl(1) | 100.57(4) | 103.65(3) |
| O(1)#2-M(1)-Cl(1) | 96.00(4) | 97.71(3) |
| N(1)-M(1)-Cl(1) | 97.76(5) | 98.79(3) |
| O(2)-M(1)-Cl(1) | 91.21(4) | 92.52(3) |
| O(1)-M(1)-Cl(1) | 174.79(4) | 174.64(3) |

SYNOPSIS TOC. Single crystal high-field electron paramagnetic resonance (HFEPR) experiments were carried out for the doped sample [Zn$_{3.91}$Ni$_{0.09}$(hmp)$_4$(dmb)$_4$Cl$_4$] in order to obtain the zero-field-splitting (zfs) parameters for the individual Ni$^{II}$ ions in the [Ni(hmp)(dmb)Cl]$_4$ cubane complex. Determination of zfs interaction at the Ni$^{II}$ sites not only provides information about the origin of the zfs parameters in the spin Hamiltonian ($D$, $E$ and $B_4^4$ etc.) for the $S = 4$ ground state of the cubane, but also provides an explanation of the fast quantum tunneling of magnetization observed for the [Ni(hmp)(dmb)Cl]$_4$ single-molecule magnet.